# Neural ODE with Temporal Convolution and Time Delay Neural Networks for Small-Footprint Keyword Spotting

*Hiroshi Fuketa[1,2] and Yukinori Morita[1]*

[1]Device Technology Research Institute    [2]AI Chip Design Open Innovation Laboratory
National Institute of Advanced Industrial Science and Technology (AIST)
E-mail: h-fuketa@aist.go.jp

## ABSTRACT

In this paper, we propose neural network models based on the neural ordinary differential equation (NODE) for small-footprint keyword spotting (KWS). We present techniques to apply NODE to KWS that make it possible to adopt Batch Normalization to NODE-based network and to reduce the number of computations during inference. Finally, we show that the number of model parameters of the proposed model is smaller by 68% than that of the conventional KWS model.

## 1. INTRODUCTION

Keyword spotting (KWS), which detects pre-defined keyword from input audio data, draws attention as a promising technique to realize so-called "voice user interface" on mobile phones, smart speakers, and so on. Recently, many researchers have demonstrated KWS with artificial neural networks (NN) and have achieved high inference accuracy [1-6]. Voice-controlled devices are, however, usually battery-operated, and hence memory footprint and compute resources are severely restricted [1]. Thus, various researches on reducing the numbers of model parameters and computations for NN-based KWS have been conducted [1,3-5].

In this paper, we propose new NN models for KWS using the neural ordinary differential equation (NODE), which was originally presented in [6]. We also propose techniques to apply Batch Normalization [7] to the NODE-based NN and to reduce the number of computations by relaxing the error tolerance of ordinary differential equation (ODE) solver. We finally show that the number of parameters of the proposed NODE-based NN is smaller by 68% than that of the conventional KWS model.

## 2. RELATION TO PRIOR WORK

Sainath and Parada [2] adopted a convolutional neural network (CNN) to KWS, which achieves better accuracy yet requires plenty of parameters. To reduce the number of parameters, various NN models have been proposed based on a residual network (ResNet) [3], temporal convolutional neural network (TCNN) [4], and time delay neural network (TDNN) [5,8]. These models consist of 5-15 stacked NN layers. In this paper, we apply NODE to KWS for the first time, which can reduce the number of layers down to 3.

R. Chen, et al. proposed NODE that interprets a residual network as ODE [6]. In this previous work, however, NODE is adopted to only simple task, such as MNIST handwritten digits recognition. It has been unknown whether applying NODE to more complicated tasks, such as KWS, is feasible. Furthermore, the increase of the number of computations caused by solving ODE is not taken into account in [6]. In this paper, we reveal that NODE can be applied to KWS by using the proposed layer-dependent batch normalization (see Section 3.2) and the proposed technique to reduce the number of computations during inference (see Section 3.3).

## 3. MODEL IMPLEMENTATION

### 3.1. Overall Model Architecture

In this paper, we propose two sorts of NN models based on the neural ODE (NODE) [6]; first one is the NODE-based network with TCNN [4] and second one is the network with TDNN [5,8]. Figure 1 illustrates the overall network architecture of the proposed models.

NODE is a novel NN algorithm proposed in [6] that interprets a residual network as discretized ODE and equivalently processes the residual network by solving the ODE. A sequence of transformation from layer $t$ to layer $t + 1$ in a residual network is given by

$$\boldsymbol{h}_{t+1} = \boldsymbol{h}_t + f(\boldsymbol{h}_t, \boldsymbol{\theta}_t), \qquad (1)$$

where $t \in \{0 \cdots N\}$ ($N$: the number of layers of residual network) and $\boldsymbol{h}_t$ is a hidden state of layer $t$. $f(\boldsymbol{h}_t, \boldsymbol{\theta}_t)$ corresponds to one layer of a residual network. Here, we introduce small layer step $\Delta t$. Assuming (1) is considered as a sequence from layer $t$ to layer $t + \Delta t$ and $\Delta t \to 0$, the following ODE is derived [9];

$$\lim_{\Delta t \to 0} \frac{\boldsymbol{h}_{t+\Delta t} - \boldsymbol{h}_t}{\Delta t} = \frac{d\boldsymbol{h}(t)}{dt} = f(\boldsymbol{h}(t), t, \boldsymbol{\theta}). \qquad (2)$$

When the input $\boldsymbol{h}(t = 0)$ is given, $\boldsymbol{h}(t = T)$, which equivalently corresponds to the output of the residual network expressed as (1), can be obtained by solving (2), where $T$ is defined as the depth of ODE solution and is treated as one of hyper-parameters in this paper. The residual network requires weight parameters of $N$ layers ($\boldsymbol{\theta}_1, \boldsymbol{\theta}_2, \cdots, \boldsymbol{\theta}_N$ in (1)), whereas NODE requires parameters of just one layer ($\boldsymbol{\theta}$ in (2)). This means the number of model parameters of NODE is much smaller than that of the residual network.

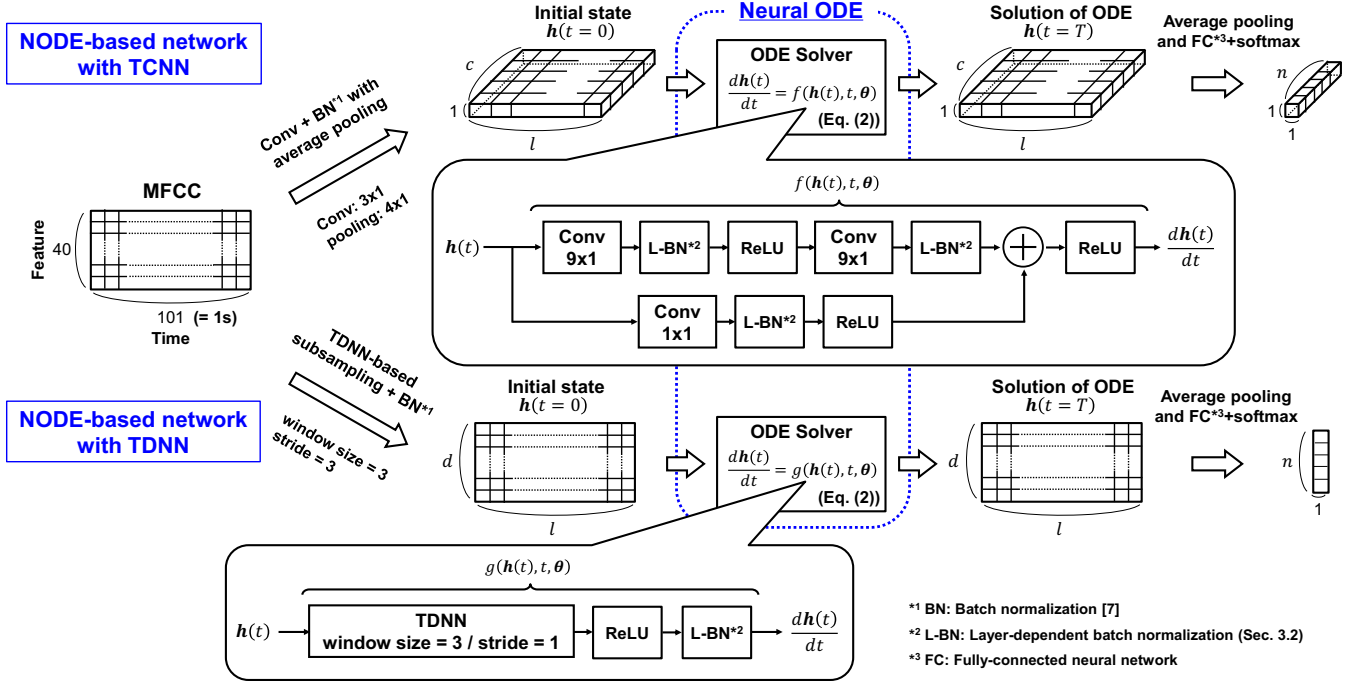

**Fig. 1.** Architecture of proposed models (NODE-based networks with TCNN and TDNN).

For both of proposed models, mel-frequency cepstrum coefficients (MFCCs) transformed from raw audio data are used as the input of the proposed networks, as shown in Figure 1. For NODE-based network with TCNN explained in the upper side of Figure 1, the MFCCs are input to the CNN layer and then to the average pooling layer. The output is 3D-matrix of $l \times 1 \times c$, where $l$ is the sequence length and $c$ is the number of channels (feature maps). This matrix is used as the initial state $\boldsymbol{h}(t=0)$ of ODE in (2). Using an ODE solver, the solution of ODE $\boldsymbol{h}(t=T)$ is obtained. It should be noted that the numbers of input and output channels in the convolutional layers ($=f(\boldsymbol{h},t,\boldsymbol{\theta})$) for ODE are set to be identical, unlike the conventional TCNN [4]. For NODE-based network with TDNN explained in the lower side of Figure 1, the MFCCs are input to the TDNN-based subsampling layer at first, which is defined in [5]. The output is $l \times d$ matrix, where $d$ is the dimensionality of feature. This matrix is used as the initial state $\boldsymbol{h}(t=0)$ of ODE containing TDNN layer ($=g(\boldsymbol{h},t,\boldsymbol{\theta})$ in Figure 1). Using an ODE solver, $\boldsymbol{h}(t=T)$ is derived.

### 3.2. Layer-Dependent Batch Normalization (L-BN)

Batch Normalization (BN) is widely used for accelerating the training of NN [7]. When BN is used in NODE, it can be expressed as

$$\mathrm{bn}(\boldsymbol{x}(t)) = \frac{x(t) - E[x(t)]}{\sqrt{V[x(t)] - \epsilon}}, \quad (3)$$

where $\boldsymbol{x}(t)$, $E[\boldsymbol{x}]$, and $V[\boldsymbol{x}]$ are a mini-batch of input data at layer $t$, mean and variance of the mini-batch, respectively. $\epsilon$ is a constant added for numerical stability ($\epsilon = 10^{-5}$ is used in this work).

In the conventional BN, the running estimates of the mean and variance of the mini-batch are calculated during training, and then they are used for BN during inference. However, this method cannot be applied to BN used in NODE, since the mean and variance depend on layer $t$, yet $t$ is a real number in NODE ($t$ is a natural number in conventional neural networks, for example, $t = 34$ in ResNet-34). Another method for the conventional BN is that the mean and variance are calculated from input data of each layer during inference. Although this method can be adopted to BN used in NODE, the inference accuracy dramatically worsens as the mini-batch size decreases, as depicted in Figure 2. For KWS, this is especially critical, since the mini-batch size during inference is usually one in practical use case.

Thus, we propose a layer-dependent batch normalization (L-BN) technique for NODE in this work. Figure 3 illustrates the brief explanation of the proposed L-BN. During training, $E[\boldsymbol{x}]$ and $V[\boldsymbol{x}]$ are calculated at each layer $t$, and then they are stored to a database. $E[\boldsymbol{x}]$ and $V[\boldsymbol{x}]$ are collected throughout one epoch. During inference, $E[x]$ and $V[x]$ are fetched from the database by using $t$ as index, and then (3) is calculated. When layer $t$ does not exist in the database, $E[x]$ and $V[x]$ are estimated by linear interpolation between data points of the nearest two layers. Figure 2 shows that the inference accuracy is not deteriorated thanks to the proposed L-BN even if the mini-batch size is 1 and the inference accuracy with the prosed L-BN is 11 times higher than that with the conventional BN.

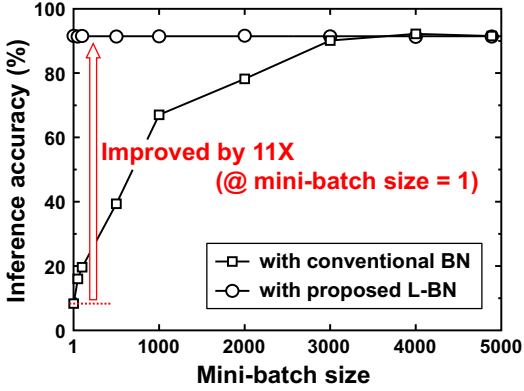

**Fig. 2.** Inference accuracy as a function of mini-batch size.

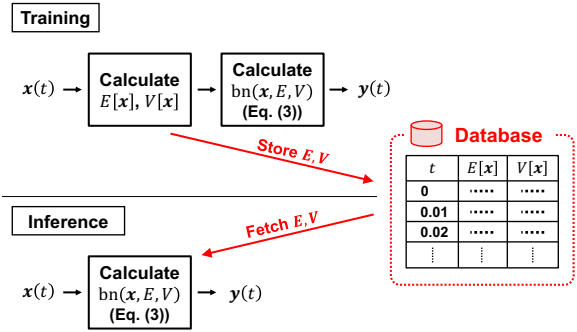

**Fig. 3.** Implementation of the proposed L-BN.

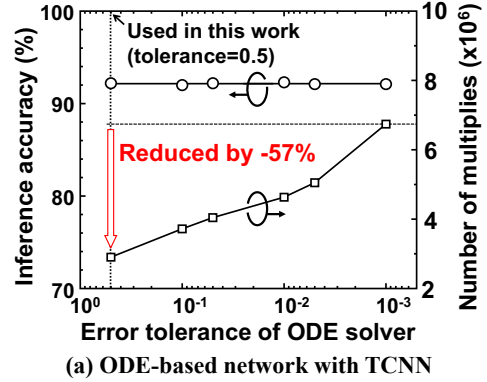

(a) ODE-based network with TCNN

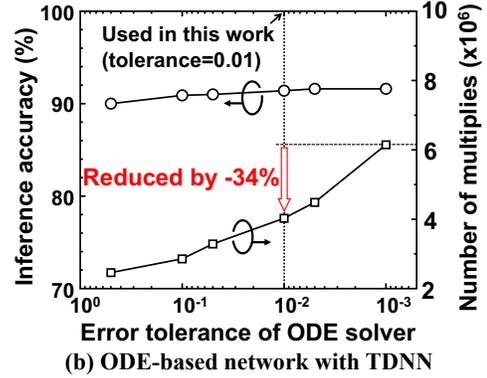

(b) ODE-based network with TDNN

**Fig. 4.** Computation reduction by tolerance relaxation.

### 3.3. Tolerance Relaxation for Inference

The error tolerance of ODE solver, which represents how accurate the solution of ODE solver is, is treated as a hyper-parameter in NODE. Although the identical tolerance values are used for training and inference in the conventional work [6], we found that the tolerance can be relaxed during inference at the very small cost of inference accuracy, which results in the dramatic reduction in the number of calculations.

Figure 4 shows the inference accuracy and the number of multiplies in the inference pass as a function of the error tolerance. The model is trained with the tolerance of $10^{-3}$. For the proposed NODE-based model with TCNN (Figure 4(a)), the tolerance during inference can be relaxed up to 0.5 and the number of multiplies can be reduce by 57% without any deterioration of accuracy, whereas for the proposed NODE with TDNN (Figure 4(b)), the tolerance during inference can be only increased to $10^{-2}$, since the accuracy slightly worsens when the tolerance is higher than $10^{-2}$. In this case, the number of multiplies in the inference pass decreases by 34%.

### 4. EXPERIMENTAL SETUP

We use Google Speech Commands Dataset [10] to evaluate the proposed models in this work. This dataset contains one-second long utterances of 30 short words. The task of our experiments follows the baseline implementations [3,4] that discriminate among 12 classes (n=12 in Figure 1) of "yes," "no," "up," "down," "left," "right," "on," "off," "stop," "go", unknown, and silence. The dataset is divided into training, validation, and test sets, whose sizes are 36921, 4443, and 4888, respectively.

In this work, we fully obey the preprocessing procedure and the feature extraction method performed in [3]. 40 MFCC features with the window size of 30 ms and the stride of 10 ms are used in this work, as shown in Figure 1. The proposed models are trained with stochastic gradient descent with momentum of 0.9, the starting learning rate of 0.1, the mini-batch size of 64, and the total epochs of 30, which corresponds to around 17k training steps. We use the Dormand-Prince (DOPRI) method as an ODE solver and the error tolerance is set to $10^{-3}$ during training. The other details about the implementations of the proposed models are described below:

---

\* Source code is available at https://github.com/fkhiro/kws-ode

**Table 1.** Setup of the proposed models.

**(a) ode-tcnn20**

| Layer | $m$ | $r$ | $c$ | $l$ | # param | # mult. | |
|---|---|---|---|---|---|---|---|
| Input | | | 40 | 101 | | | |
| Conv | 3 | 1 | 20 | 101 | 2.4k | 242k | |
| Avg. pool | 4 | 1 | 20 | 25 | | | |
| Conv | 9 | 1 | 20 | 25 | 3.6k | 90k | |
| Conv | 9 | 1 | 20 | 25 | 3.6k | 90k | ODE |
| Conv | 1 | 1 | 20 | 25 | 0.4k | 10k | |
| Avg. pool | | | 20 | 1 | | | |
| FC | | | 12 | 1 | 0.24k | 240 | |
| Total | | | | | 10k | 242k + 190k × $NFE^*$ | |

**(b) ode-tdnn32**

| Layer | $w$ | $s$ | $d$ | $l$ | # param | # mult. | |
|---|---|---|---|---|---|---|---|
| Input | | | 40 | 101 | | | |
| TDNN-SUB | 3 | 3 | 32 | 34 | 3.9k | 131k | |
| TDNN | 3 | 1 | 32 | 34 | 3.1k | 104k | ODE |
| Avg. pool | | | 32 | 1 | | | |
| FC | | | 12 | 1 | 0.4k | 384 | |
| Total | | | | | 7.4k | 131k + 104k × $NFE^*$ | |

\* NFE: The number of function evaluations [6]

**Table. 2.** Comparison of proposed models to baselines.

| Model | Accuracy (%) | # param |
|---|---|---|
| trad-fpool3 [2]* | 90.5 | 1.4M |
| tpool2 [2]* | 91.7 | 1.1M |
| res8-narrow [3]* | 90.1 | 20k |
| res15-narrow [3]* | 94.0 | 43k |
| res15 [4]* | 95.8 | 240k |
| tc-resnet8 [4]* | 96.1 | 66k |
| tc-resnet14-1.5 [4]* | 96.6 | 310k |
| tdnn [5]*,¶ | 94.4 | 12k |
| swsa [5]*,¶ | 90.2 | 8k |
| ode-tcnn30 | 93.6 | 21k |
| ode-tcnn20 | 92.2 | 10k |
| ode-tdnn32 | 91.4 | 7.4k |
| ode-tdnn29 | 90.4 | 6.4k ✓ |

(This work: ode-tcnn30, ode-tcnn20, ode-tdnn32, ode-tdnn29)

\* Accuracy is taken from the paper.

¶ Listed just for reference and not compared with the proposed models in this paper, since they only distinguish 11 classes while our models distinguish 12 classes.

**With TCNN:** The network architecture is described in the upper side of Figure 1. We evaluate two variants referred as ode-tcnn20 and ode-tcnn30 in this paper. The setup of ode-tcnn20 is listed in Table 1(a). $m$ and $r$ are the width and height of the kernel (filter) of the convolutional layers, respectively, and the stride of the kernel is 1. $c$ is the number of channels, and $l$ is the length of the output ($c$ and $l$ are also depicted in Figure 1). For ode-tcnn30, the numbers of channels of the convolutional layers $c$ are 30 and other parameters are same as those of ode-tcnn20. During training, the learning rate is multiplied by 0.1 when the time step reaches 5k and 9k, $L_2$ weight decay is $10^{-3}$, and the depth of ODE solution $T$ is 1. During inference, the error tolerance is relaxed to 0.5, as shown in Figure 4(a).

**With TDNN:** The network architecture is described in the lower side of Figure 1. We evaluate two variants referred as ode-tdnn32 and ode-tdnn29 in this paper. The setup of ode-tdnn32 is listed in Table 1(b). $w$ and $s$ are the length and stride of TDNN window [5], respectively. $d$ and $l$ is the dimensionality and length of the output, respectively, which are depicted in Figure 1. For ode-tdnn29, the dimensionalities of TDNN-SUB and TDNN layers are 29 and other parameters are same as those of ode-tdnn32. During training, the learning rate is multiplied by 0.1 when the time step reaches 6k and 10k, $L_2$ weight decay is $10^{-5}$, and the depth of ODE solution $T$ is 3. During inference, the error tolerance is relaxed to $10^{-2}$ for ode-tdnn32 and $5 \times 10^{-3}$ for ode-tdnn29, respectively.

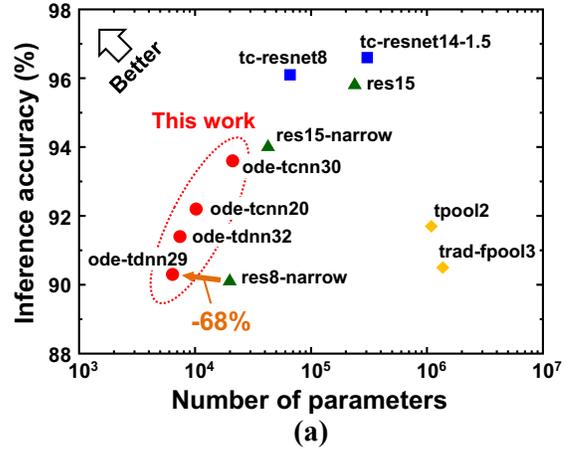

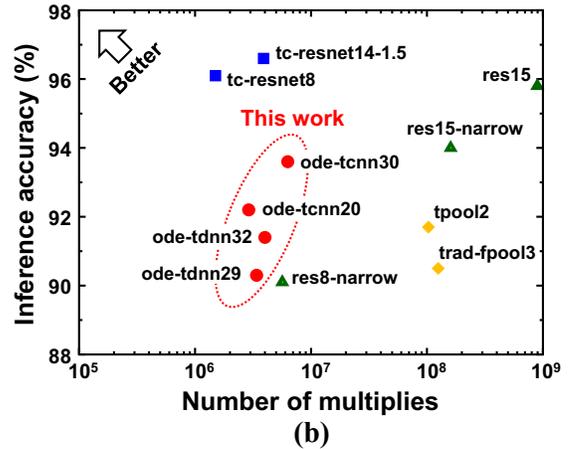

**Fig. 5.** Inference accuracy as a function of the number of (a) model parameters and (b) multiplies in the inference pass.

## 5. RESULTS

The inference accuracy and the number of weight parameters are summarized in Table 2. In this paper, we compare the proposed models with the following baseline models; 1) CNN-based KWS models have been originally proposed in [2]. They are denoted as trad-fpool3 and tpool2, whose inference accuracy is taken from [3]. 2) Tang and Lin [3] applied deep residual network (ResNet) architecture to KWS models referred as res8-narrow, res15-narrow, and res15. 3) Choi et al. [4] proposed the temporal convolutional neural network (TCNN) for KWS. We compare its two variants represented by tc-resnet8 and tc-resent14-1.5. 4) The KWS models based on the time delay neural network (TDNN) were proposed in [5] and referred as tdnn and swsa in Table 2.

Figures 5(a) and (b) show the inference accuracy of the baseline and proposed KWS models as a function of the number of model parameters and the number of multiplies in the inference pass, respectively. The proposed NODE-based model with TDNN require smaller parameters, whereas their accuracy is worse compared to the NODE-based model with TCNN. The number of parameters of the proposed model (ode-tdnn29) is smaller by 68% than that of the conventional model (res8-narrow) at iso-accuracy. The accuracy of the proposed model (ode-tcnn30) is improved by 3.5% compared to the baseline model (res8-narrow) under the same condition of the number of parameters (around 20k). In contrast, the numbers of multiplies of the proposed models are smaller than those of the conventional CNN- and ResNet-based models, whereas they are comparable to those of the TCNN-based baseline models even though the proposed tolerance relaxation technique described in Sec. 3.3 is used. This is because many cycles of calculation of the neural network, which is denoted by *NFE* in Table 1, are required to solve ODE using Runge–Kutta method in the proposed models. To tackle with this issue, a hardware acceleration of ODE solver is expected. This is future work.

## 6. CONCLUSION

In this paper, we proposed NODE-based NN models for small-footprint KWS. To adopt NODE-based NN models to KWS, we proposed L-BN to apply Batch Normalization for NODE models. Furthermore, we also optimized the error tolerance of ODE solver to reduce the number of computations during inference. Finally, we showed that the number of model parameters of the proposed model can be reduced by 68% compared to that of the conventional KWS model.

## 7. ACKNOWLEDGEMENTS

This work was partly supported by JSPS KAKENHI Grant Number JP20K11740.